\documentstyle[pramana,epsf,floats]{ias}
\begin{document}
\title{Relativistic Calculations of Coalescing Binary Neutron Stars}
\author{Joshua Faber} 
\address{Department of Physics and Astronomy, Northwestern
University}
\author{Philippe Grandcl\'ement}
\address{Laboratoire de Math\'ematiques et de Physique
 Th\'eorique, Universit\'e de Tours, Parc de Grandmont, 37200 Tours,
 France}
\author{Frederic Rasio}
\address{Department of Physics and Astronomy, Northwestern
University}

\abstract{
We have designed and tested a new relativistic Lagrangian hydrodynamics code,
which treats gravity in the conformally flat approximation to general
relativity. We have tested the resulting code
extensively, finding that it performs well 
for calculations of equilibrium single-star models,
collapsing relativistic dust
clouds, and quasi-circular orbits of equilibrium solutions.
By adding in a radiation reaction treatment, we compute the full
evolution of a coalescing binary neutron star system.  We find that
the amount of mass ejected from the system, much less than a percent,
is greatly reduced by the
inclusion of relativistic gravitation.  The gravity wave energy
spectrum shows a clear divergence away from the Newtonian point-mass
form, consistent with the form derived from relativistic
quasi-equilibrium fluid sequences.}
\maketitle

\section{Introduction}

It has long been recognized that coalescing binary neutron star (NS)
systems are a leading candidate to be the first observed source of
gravity waves (GW).  With LIGO, GEO, and TAMA all taking scientific
data, and VIRGO in the commissioning stage, it is growing increasingly
important to have quantitatively accurate predictions of the GW
signals we expect to measure during the merger process.  Besides
their use in aiding detections, these predictions are crucial for
determining important physical information about the mass, radius, and
equation of state (EOS) of NS from GW observations.

Calculations of binary NS coalescence have been performed for many
years, beginning with studies in  Newtonian gravity.  It was
recognized all along, however, that general relativity (GR) will play
an important role during the merger, since the characteristic
gravitational fields and velocities are squarely within the
relativistic regime.
As a result, increasingly sophisticated gravitational formalisms have
been used in hydrodynamical calculations, starting with  
post-Newtonian treatments \cite{FR1,FR2,FR3,APODR,SNO}, many of which
were based on a formalism developed by Blanchet et al. \cite{BDS}
which includes all lowest-order 1PN effects as well as lowest order
dissipative effects from gravitational radiation reaction losses. 
More recently,
calculations have been performed in full general
relativity \cite{SU1,SU2,STU}.  
Unfortunately, the PN approximation breaks down
during the merger when higher-order relativistic effects grow
significant, and fully relativistic calculations typically introduce
numerical instabilities which limit the amount of time for which a
calculation will remain accurate.  A middle ground is provided by the
conformally flat (CF) approximation, developed originally by Wilson et
al. \cite{WMM}, which
includes much of the non-linearity inherent in GR, but results in a
set of coupled, non-linear, elliptic field equations, which can be
evolved stably.  We assume that the spatial part of the GR
metric is equal to the flat-space form, multiplied by a
conformal factor which varies with space and time, the metric taking the
form
\begin{equation}
ds^2=-(N^2-B_i B^i)dt^2-2B_i dt dx^i+A^2 \delta_{ij}dx^i dx^j.
\end{equation}
While this approach cannot reproduce the exact GR solution for a
general matter configuration, 
it is exact for spherically symmetric systems, and yields solutions
which agree with those calculated using
full GR to within a few percent for many systems of
interest \cite{MMW}.

\section{Conformally Flat SPH}

After \cite{Baum} calculated the evolution of coalescing NS binaries
using a PN variant on the CF formalism, \cite{Oech} performed the
first dynamical calculations which included all the non-linear effects
present in the CF formalism.  
Unfortunately, the approach used in these efforts and many others
throughout the history of this line of research is not particularly
efficient. Solving the partial differential equations describing
metric fields on large grids is very costly, both in terms of time and
computer memory.  Motivated by this conclusion, we combined our
previous work in 3-d hydrodynamics with a spherical coordinates 
spectral methods code, which
decomposes all field and hydrodynamical quantities into radial and
angular functions.  Our 
field solver, based on {\tt LORENE} \cite{Gour}, is extremely efficient. 
The numerical libraries are publicly available at {\tt http://lorene.obspm.fr},
and  have been used previously, among other things, to construct
the quasi-equilibrium binary
configurations used in the aforementioned relativistic
calculations. 
We combined this field solver with a smoothed particle hydrodynamics (SPH)
evolution treatment, resulting in a 3-d code which can compute
the full evolution of any number of relativistic matter configurations
accurately and efficiently.  Our code is, as best we know, the first
3-d hydrodynamic evolution treatment of binary NS systems to use
either spectral methods or spherical coordinates.  For extremely
detailed results, please refer to \cite{FGR}, which we will refer to
hereafter as FGR.

\begin{figure}[tbp]
\epsfxsize=7cm
\centerline{\epsfbox{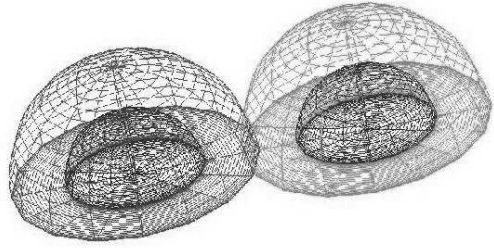}}
\caption{Radial domains used to solve the field equations of the CF
method.  Each vertex is a collocation point.  {\it From
  \protect\cite{Gour}}.}
\label{fig:grid}
\end{figure}

The field solver works by breaking up source terms into two distinct
components, each centered on a star, which are further broken into
radial domains, as shown in Fig.~\ref{fig:grid}.  
In each 
domain, terms are evaluated at ``collocation points'' spaced out in the
radial and angular directions to handle any convex surface.
Typically, solutions accurate to one part in $10^9$ can be achieved
quickly, using only a $17\times 13\times 12$ grid.
The Lagrangian nature of SPH has several advantages over Eulerian
grid-based methods for these calculations, first and foremost the
natural way it handles a surface; there are no particles where there
is no matter.

\section{Code tests}

We have performed several tests to ensure that our code works
properly.  Since the CF formalism is known to be exact for spherically
symmetric matter configurations, we calculated models of isolated
neutron stars, finding excellent agreement with the well known
Oppenheimer-Volkov solution to well within a percent for all
hydrodynamic expressions and field values throughout the star.

To test the dynamical aspects of the code, we also computed the
collapse of a dust cloud, i.e. pressureless matter, placed initially
at rest.  We compared our results to those of \cite{Petrich}, who
developed a semi-analytic procedure which yields the field values at
all points in spacetime, as well as the paths traced out by any given
mass shell.  We find that our code can reproduce the collapse
extremely well, until just short of the point where the event horizon
reaches the surface of the matter. 

Since the CF formalism is time-symmetric, it does not contain terms
which lead to gravitational radiation back reaction. Thus, we have
tested our code by computing the evolution of 
quasi-equilibrium binaries, taken from the ``M14~vs.~14'' sequence of
\cite{Tani}.  We found that 
the binary separation and conserved system angular
momentum vary by no more than $2.5\%$ over two orbits, and the ADM mass
is nearly constant for runs started at three different initial
separations, including the innermost stable value found before a cusp forms.
Similarly, a comparison of the field values and density profiles of the
stars after two orbits yields very little deviation from the initial
configuration.  
These
results confirm for the first time that equilibrium binary
configurations calculated by \cite{Tani} are dynamically stable all
the way to the appearance of a cusp.

\section{Binary NS mergers}
 
Dissipative effects can be added to the CF formalism through a
radiation reaction
potential which reproduces the lowest-order energy loss rate \cite{WMM}.
When radiation reaction is included, we find that the binary plunges rapidly
toward merger soon after passing the point where a cusp is reached along the
quasi-equilibrium sequence.  In our approach, we have
found that throughout the evolution, the NS surfaces can be
modeled by triaxial ellipsoids that are
allowed to rotate to match the growing tidal lag angles.
Field quantities are calculated by finding the SPH values
for source terms at collocation points, and solving the field
equations in the spectral basis.
Field values are then interpolated back to SPH particles, with
derivatives calculated to high accuracy by the field solver, rather
than particle-based techniques.  For overlapping configurations, we
split our source terms between the two stars, weighting the density
contributions such that each NS has a well-defined central
density maximum, up until the point where  the central density of the
system allows us to treat the object as a single rapidly spinning body. 

\begin{figure}[tbp]
\epsfxsize=7.5cm
\centerline{\epsfbox{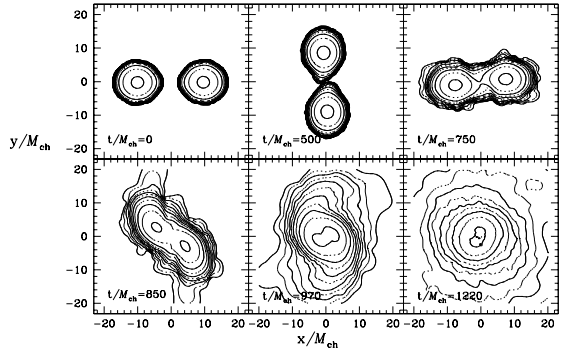}}
\caption{The evolution of the coalescing NS binary.
  Density contours are logarithmically spaced, two per
  decade. We see the development of significant tidal lag angles at $t/{\cal
  M}_{ch}=500$, followed by an ``off-center'' collision that 
leads to the formation of a vortex sheet
  and a small amount of matter ejection, finally resulting in a
  differentially rotating spheroidal configuration which is stable
  against gravitational collapse.  Units are defined such that
  $G=c=1$, and ${\cal M}_{ch}$ is the chirp mass of the system.} 
\label{fig:dens}
\end{figure}

The evolution of the NS during a coalescence is shown in
Fig.~\ref{fig:dens}.  
In turn we see the inspiral of the NS, which
orbit counterclockwise, with tidal lags growing as they do so.  When
the NS make contact, they collide in an ``off-axis'' manner, with a
very small amount of mass running along the surface interface before
being spun off the newly forming remnant.  This trace amount of
matter, representing much less than $1\%$ of the total system mass,
remains gravitationally bound, forming a tenuous halo around the
rapidly and differentially rotating ``hypermassive'' remnant.

\begin{figure}[b]
\epsfxsize=7.0cm
\centerline{\epsfbox{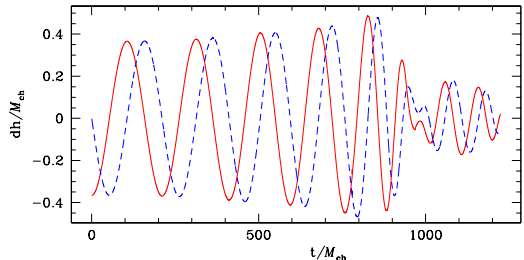}}
\caption{Gravitational wave signal in the $h_+$ (solid line) and
  $h_\times$ (dashed line) polarizations, for an observer located a
  distance $d$ from the system along the rotation axis.  Units are as
  in Fig.~\protect\ref{fig:dens}.  We see a chirp signal during the
  inspiral, followed by a lower-amplitude, modulated 
burst of high-frequency emission
  while the remnant forms.}
\label{fig:gwpl}
\end{figure}

We calculate the GW signal produced during the merger in the
lowest-order quadrupole limit, finding good agreement between our
results and the relativistic calculations of \cite{STU}.  In
Fig.~\ref{fig:gwpl}, we show the gravity wave signal in both
polarizations, $h_+$ and $h_\times$, as a function of time.
Prior to the merger, we see a ``chirp'' signal, as the frequency and
amplitude both increase while the stars approach each other.  After a
remnant forms, there is a period of modulated high-frequency
emission, which damps away as the remnant relaxes toward a spheroidal
shape.  

While the time-dependence of the GW signal is important, it is
perhaps more enlightening to look at the frequency dependence of the
signal, and in particular, the energy spectrum $dE_{GW}/df$.  
Indeed, we have argued previously \cite{FGRT} that the changes in
the total energy of quasi-equilibrium binary configurations for NS
models with different compactness values $M/R$  should
leave an imprint in the energy spectrum in the form of a ``break''.
This can be observed by a
narrow-band detector on an advanced interferometer such as LIGO II.
Our argument is extremely straightforward: for a given sequence, we
can calculate the total energy and GW frequency as a function of the
binary separation, finding in general that the former is sensitive to
the compactness of the NS, while the latter is not.  After implicitly
determining $E(f)$, we can numerically differentiate with respect to
the GW frequency to find the energy spectrum. Relativistic  effects
typically flatten out the equilibrium energy curve with respect to the
binary separation, which 
decreases the amplitude of the 
energy spectrum at the corresponding frequency. Using a
parameterized model of these ``break frequencies'',
Hughes \cite{Hughes}
determined that the NS radius could be determined to within a few
percent with at most $\sim 50$ LIGO II observations of coalescing NS,
and perhaps far fewer for optimal parameter values.

In order to avoid aliasing when taking the Fourier transform of the GW
signal, we need to attach some estimate of the signal behavior both
before and after the period which we calculate.  Noting that the
frequencies corresponding to remnant emission may be impossible to
detect even with LIGO II, we fit the portion after our calculation with
an exponentially damped oscillatory signal.  The inspiral signal is
much more important, but most
groups have traditionally fit the inspiral by the lowest-order
``Newtonian'' point-mass form \cite{FR3,Oech}.  This approach can lead
to qualitatively inaccurate results, however, because it does not
account for finite-size and relativistic corrections.    Noting
this, in FGR we calculated an inspiral waveform directly from the
quasi-equilibrium sequence which we used to construct our initial
configuration.  The inspiral waveform is thus completely consistent
with our calculation.  

\begin{figure}[tbp]
\epsfxsize=7.5cm
\centerline{\epsfbox{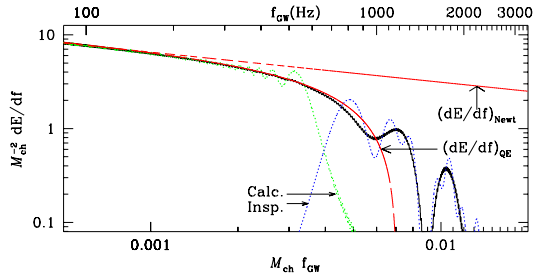}}
\caption{GW energy spectrum, ${\cal M}_{ch}^{-2}dE/df$, as a
  function of the GW frequency, ${\cal M}_{ch} f_{GW}$.
  The dotted lines show, respectively at high and low
  frequencies, the components contributed by our calculated signal
  and the quasi-equilibrium inspiral component.  Also shown are the Newtonian
  point mass energy spectrum (short-dashed line), 
  and the quasiequilibrium fit derived from
  equilibrium sequence data.  On the
  upper axis, we show the corresponding frequencies in Hz assuming the
  NS each have an ADM mass $M_0=1.4M_\odot$.  We see that the ``break frequency''
  occurs well within the sub-kHz regime.}
\label{fig:rr_gwps}
\end{figure}

Since our calculation was started from a quasi-circular orbit, we
attach the inspiral waveform onto our calculated merger waveform at
the point where the binary had attained the proper infall velocity,
which we determined to be at $t=250$.
We note, however, that the exact point where the crossover was made
has virtually no effect on the resulting spectrum.
In Fig.~\ref{fig:rr_gwps}, we show as a solid line
a complete and {\it consistent} relativistic waveform for
a binary NS merger.  The frequencies listed on the upper axis assume typical
parameters for NS: each has an ADM mass $M_0=1.4
M_{\odot}$.  
The two dotted lines show the components which make up
the energy spectrum.
At low frequencies, the primary contribution is from the inspiral
waveforms, and at high frequencies, from the calculated merger
waveform.  The short-dashed line shows the Newtonian point-mass
relation, $(dE/df_{GW})_N=\pi^{2/3}{\cal
  M}_{ch}^{5/3}f_{GW}^{-1/3}/3$, and the long-dashed curve the fit we
find from our quasi-equilibrium sequence data.  
 We see excellent agreement between our calculated
waveform and the quasi-equilibrium fit, up until frequencies ${\cal
  M}_{ch}f_{GW}\approx 0.007-0.009$.  This peak represents the
``piling up'' of energy at the frequency corresponding to the phase of
maximum GW luminosity, as the stars make contact and the infall rate
drops dramatically.  The second peak, at ${\cal
  M}_{ch}f_{GW}\approx 0.010-0.011$, represents emission from the
ringdown of the merger remnant.  It is likely that we underestimate
the true height of this second peak somewhat, since we assume that the GW
signal after our calculation damps away exponentially, but even for
higher-amplitude peaks detections
at these frequencies would be nearly impossible anyway.  

In general, the energy spectrum we calculated confirms the general
conclusions put forward in \cite{FR3,FGRT},
albeit in a much more consistent way.  The GW energy spectrum differs
from the Newtonian point-mass form at frequencies much less than
$1~{\rm kHz}$, within the range accessible to LIGO II..
Thus, combining sub-kHz narrow-band detectors
with broadband LIGO measurements, as suggested in \cite{Hughes},
should allow GW measurements to constrain the NS compactness and EOS.  

This work was supported by NSF grants PHY-0133425 and PHY-0245028 to
Northwestern University.

\end{document}